\documentclass[twocolumn,tighten]{aastex631}
\usepackage{amsfonts,amsmath,amssymb}
\usepackage{bm}
\usepackage{xcolor}
\usepackage[]{cleveref}
\crefname{section}{\S\!}{\S\S\!}
\crefname{appendix}{\S\!}{\S\S\!}
\crefname{equation}{Eq.}{Eqs.}
\Crefname{equation}{Equation}{Equations}
\crefname{figure}{Fig.}{Figs.}
\Crefname{figure}{Figure}{Figures}
\usepackage{mathtools,upgreek,pifont}
\usepackage{cuted}

\newcommand{\ffold}{}
\newcommand{\ta}{\tau_{\rm A}}
\newcommand{\va}{v_{\rm A}}

\newcommand{\rmd}{{\rm d}}

\newcommand{\pD}[2]{\frac{\partial #2}{\partial #1}}

\newcommand{\D}[2]{\frac{\rmd #2}{\rmd #1}}

\newcommand\bb[1]{\mbox{\boldmath{$#1$}}}
\newcommand\grad{\bb{\nabla}}
\newcommand\bcdot{\,\bb{\cdot}\,}

\newcommand\btimes{\,\bb{\times}\,}

\newcommand{\ez}{\hat{\bb{z}}}

\newcommand{\nathb}{S+22}

\submitjournal{Astrophys. J. Lett.}

\shorttitle{Electron-ion heating partition}
\shortauthors{Squire et al.}

\begin{document}

\title{Electron-ion heating partition in imbalanced solar-wind turbulence}

\author[0000-0001-8479-962X]{Jonathan Squire}
\affiliation{Physics Department, University of Otago, Dunedin 9010, New Zealand}
\email{jonathan.squire@otago.ac.nz}

\author[0000-0002-8327-5848]{Romain Meyrand}
\affiliation{Physics Department, University of Otago, Dunedin 9010, New Zealand}

\author[0000-0003-1676-6126]{Matthew W.~Kunz}
\affiliation{Department of Astrophysical Sciences, Princeton University, Peyton Hall, Princeton, NJ 08544, USA}
\affiliation{Princeton Plasma Physics Laboratory, PO Box 451, Princeton, NJ 08543, USA}

\begin{abstract}
A likely candidate mechanism to heat the solar corona and solar wind is low-frequency ``Alfv\'enic'' turbulence sourced by magnetic fluctuations
near the solar surface. Depending on its properties, such turbulence can heat different species via different mechanisms, and 
the comparison of theoretical predictions to observed temperatures, wind speeds, anisotropies, and their variation with heliocentric radius provides
a sensitive test of this physics. Here  we explore the importance of normalized cross helicity, or imbalance, for controlling solar-wind heating, 
since it is a key parameter of magnetized turbulence and varies systematically with wind speed and radius.
Based on a hybrid-kinetic simulation in which the forcing's imbalance decreases with time---a crude model for a plasma parcel  entrained in the outflowing wind---we demonstrate how significant changes to the turbulence and heating result from the ``helicity barrier'' effect.
Its dissolution at low imbalance causes its characteristic features---strong perpendicular ion heating with a steep ``transition-range'' drop in electromagnetic fluctuation spectra---to disappear, 
driving more energy into electrons and parallel ion heat, and halting the emission of ion-scale waves. These predictions seem to agree
with a diverse array of solar-wind observations, offering to explain a variety of complex correlations and features within a single theoretical framework.
\end{abstract}

\keywords{}

\section{Introduction} \label{sec:intro}

The solar corona and its extended outflow, the solar wind, provide us with 
an unparalleled laboratory for studying the physics of magnetized collisionless plasmas. Decades of observations have
revealed a highly complex system filled with electromagnetic fluctuations across a vast range of scales, whose properties (e.g., total power or spectra)
are correlated in surprising and nontrivial ways with the plasma's flow speed, temperatures, anisotropies, and elemental abundances \citep{Marsch2006,Horbury2012,Bruno2013}. 
These correlations, as well as the extended heating at large heliocentric distances needed to explain the high speed of fast-wind streams \citep{Parker1965},
suggest that the solar wind is shaped by both properties of its low-coronal source and turbulent heating at larger altitudes. 
Of particular interest are the decades 
of observations that hint at the role played by the fluctuations' \emph{imbalance} (i.e., normalized cross helicity), 
 a key parameter in the theory of  magnetized turbulence
\citep{Dobrowolny1980a,Schekochihin2020} that is observed to correlate with wind speed $U$ and 
decrease with increasing heliocentric 
radius $R$ \citep[e.g.,][]{Roberts1987,Marsch2006,DAmicis2021}.

In this Letter, we argue for the importance of imbalance in shaping turbulence and heating in the low-$\beta$ solar wind. 
We focus on the physics of the ``helicity barrier'' \citep{Meyrand2021}, which dramatically alters imbalanced turbulence in $\beta\lesssim1$
plasmas by restricting the turbulent energy cascade at perpendicular scales $\lambda_{\perp}$ below the ion gyroradius $\rho_{i}$.
Previous kinetic simulations have shown how numerous features of helicity-barrier-mediated turbulence match measurements of the low-$\beta$ solar wind  well (\citealp{Squire2022}; hereafter \nathb), 
including properties of the steep ``transition-range'' drop  in electromagnetic field spectra around $\rho_{i}$ scales  and the ion velocity distribution function (VDF; see, e.g., \citealp{Duan2021,Bowen2022,Bowen2023}).

We study how the turbulence evolves as 
 imbalance is slowly decreased in time, transitioning between the highly imbalanced and balanced regimes. 
This physics is of interest for two reasons: first, it is a crude 
model for the expected response of the turbulent heating as the plasma flies outwards from the Sun and its imbalance decreases;  second, 
it captures how the helicity barrier becomes weaker and eventually disappears at low imbalance,  thus probing its robustness. 
The results, which are based on a kinetic simulation of forced Alfv\'enic turbulence, can explain the observed correlation of wind speed with ion temperatures, as well as the switch from negative to positive correlation of wind speed with electron temperature at increasing $R$ \citep[e.g.,][]{Burlaga1973,Marsch1989,Shi2023,Bandyopadhyay2023}. 
Correlations with other properties such as the  spectral transition range, proton VDFs, and plasma-wave/instability activity seem
to explain a diverse array of observations within a single theoretical framework.

\section{Formulation of the problem and method of solution}

\subsection{The helicity barrier}\label{sub: helicity barrier}

The helicity barrier arises due to the conservation of a ``generalized helicity'' $\mathcal{H}$ in low-$\beta$ gyrokinetics \citep{Schekochihin2019,Meyrand2021}. At large scales $\lambda_{\perp}\gg \rho_{i}$, 
$\mathcal{H}$ is the cross helicity, which, along with the energy, cascades forward  to smaller scales; at small scales $\lambda_{\perp}\ll \rho_{i}$, $\mathcal{H}$ is a form of magnetic
helicity, which cannot cascade to smaller $\lambda_{\perp}$ without violating energy conservation \citep{Fjortoft1953}. Thus, the sub-$\rho_{i}$  cascade 
must have $\mathcal{H}=0$, restricting its energy flux to be at most the ``balanced portion'' of that injected, $\varepsilon-\varepsilon_{H}$, where $\varepsilon$ and $\varepsilon_{H}$ are the large-scale injection rates of energy  and helicity, respectively. The remaining energy input, $\varepsilon-(\varepsilon-\varepsilon_{H})=\varepsilon_{H}$, is
``trapped'' at $\lambda_{\perp }\lesssim \rho_{i}$ scales, causing the turbulence to reach large amplitudes and (through critical balance)  small parallel scales approaching the ion inertial length $d_{i}$.   The fluctuations then take on the character of
 oblique ion-cyclotron waves (ICWs) with frequencies $\omega$ that approach the ion gyrofrequency $\Omega_i$. This
 activates strong wave-particle interactions and quasi-linear resonant heating \citep{Kennel1966}, which can absorb the remaining energy input, allowing the turbulence to saturate \citep{Li1999}.  It also causes the plasma to become unstable and emit small-scale parallel ICWs \citep{Chandran2010a}.
 By diverting a large fraction of the energy flux that would have reached electron scales, and thus electron heat, into ions instead, the
 helicity barrier causes large ion-to-electron heating ratios $Q_{i}/Q_{e}\simeq\varepsilon_{H}/(\varepsilon-\varepsilon_{H})$ in anisotropic low-$\beta$ turbulence.

\subsection{Numerical method}\label{sec: numerics}

We use the hybrid-kinetic method implemented in the \texttt{Pegasus++} code \citep{Kunz2014a,Arzamasskiy2022}. This approach 
treats the ion (proton) dynamics fully kinetically using a particle-in-cell (PIC) approach, while the electrons constitute a massless, neutralizing, isothermal fluid. The ion macro-particle positions ($\bm{r}$) and velocities ($\bm{v}$) are drawn from an initially Maxwellian $f_i(\bb{v})$ and evolved via
\begin{equation}
\D{t}{\bm{r}}= \bm{v},\quad \D{t}{\bm{v}}=\frac{e}{m_{i}}\left[\bm{E}(\bm{r},t) + \frac{\bm{v}}{c}\btimes \bm{B}(\bm{r},t)\right] + \frac{1}{m_{i}}\bm{F}_{\perp}^{U},
\end{equation}
where $\bb{E}$ and $\bb{B}$ are the electric and magnetic fields and $\bm{F}_{\perp}^{U}$ stirs turbulence by injecting incompressible motions perpendicular (``$\perp$'') to a mean (``guide'') magnetic field $\bm{B}_0$. 
The magnetic field satisfies a modified version of Faraday's law,
\begin{equation}
\pD{t}{\bm{B}}= -c\grad\btimes (\bm{E} + \bm{F}_{\perp}^{B}) +\eta_{4} \nabla^{4}\bm{B},
\end{equation}
where $\bm{F}_{\perp}^{B}$ forces solenoidal magnetic fluctuations perpendicular to $\bm{B}_0$ and $\eta_{4}$ is a hyper-resistivity that absorbs small-scale magnetic energy. The electric field is
\begin{equation}
\bm{E} = - \frac{\bm{u}}{c}\btimes \bm{B} - \frac{T_{e}}{e n} \grad{n} + \frac{(\grad\btimes \bm{B})\btimes \bm{B}}{4\pi e n},
\end{equation}
where $n$ is the ion (and electron) density and $\bm{u}$ is the ion flow velocity, both of which are computed via a weighted sum of the macro-particles in the relevant region of space; $T_{e}$ is the electron temperature, which is a parameter of the model; and $e$, $m_{i}$, and $c$
are the electron/ion charge, the ion mass, and the speed of light, respectively. We also define the Alfv\'en speed $\va\equiv B_{0}/\sqrt{4\pi n m_{i}}$, $\Omega_{i}\equiv eB_{0}/(m_{i}c)$,  the ion thermal speed $v_{\rm th}$, and the ``Els\"asser'' fields $\bm{z}^{\pm}\equiv\bm{u}_{\perp}\pm\bm{B}_{\perp}/\sqrt{4\pi n m_{i}}\equiv \bm{u}_\perp\pm \bm{b}_\perp$. Volume averages are denoted by $\langle\,\cdot\,\rangle$ and $f_{i}(w_{\perp},w_{\|})$ is the gyro-averaged VDF, with $w_{\perp}$ and $w_{\|}$  the particle velocities perpendicular and
parallel to the local magnetic field in the frame of the plasma (i.e., $\bm{w}\equiv\bm{v}-\bm{u}$).

\subsection{Problem set up}

Our basic simulation set up follows \nathb. The simulation represents a small co-moving patch of plasma, capturing realistic solar-wind
turbulence amplitudes and anisotropies around $k_{\perp}\rho_{i}\sim1$ scales, where $k_{\perp}\sim1/\lambda_{\perp}$ denotes the perpendicular wavenumber. 
The  domain is Cartesian and periodic, with coordinates $\{x,y,z\}$,
and is elongated along $\bm{B}_{0}=-B_{0}\ez$, with $L_{z}=6L_{\perp}$ and $L_{\perp}=67.5d_{i}$ ($d_{i}=\va/\Omega_{i}$).
With these parameters, 
the box shape  approximately matches the  (statistical) shape of turbulent eddies measured at similar scales in the solar wind (\citealp{Chen2016}; \nathb).
The grid resolution is $N_{\perp}^{2}\times N_{z}=392^{2}\times 2352$, so that the 
smallest resolved scales are $k_{\perp,{\rm max}}d_{i}\simeq\pi N_{\perp}d_{i}/L_{\perp}\approx 18$.  
We use $N_{\rm ppc}=216$ ion-macro-particles per cell. The hyper-resistivity was increased  from $\eta_{4}\approx2.4\times10^{-5}d_{i}^{4}\Omega_{i}$  to
 $\eta_{4}=5\times10^{-5}d_{i}^{4}\Omega_{i}$ at $t=4.3\ta$ in response to the strengthening kinetic-range cascade.

The simulation is initialized at $t=0$ using  the final snapshot from \nathb, and thus represents saturated
highly imbalanced turbulence with strong perpendicular ion heating and a helicity barrier, similar to turbulence observed in the near-Sun solar wind by PSP \citep{Bowen2023}.
The initial ion temperature is such that $\beta_{i}=8\pi\langle nT_{i}\rangle/\langle B^{2}\rangle\approx0.33$; electrons are isothermal with $\beta_{e0}=8\pi\langle n\rangle T_{e}/B_{0}^{2}=0.3$. The simulation 
is run for an additional ${\approx}18\ta$, where $\ta=L_{z}/\va$ is the outer-scale Alfv\'en time, which is also comparable to the turnover time of the turbulence $\tau_{\rm turb}\sim L_{\perp}/u_{{\rm rms}}$ due to our choice of forcing parameters (see below; $u_{{\rm rms}}=\langle u^{2}\rangle^{1/2}$ is the root-mean-square velocity). 
Because the ions heat up, $\beta_{i}$ increases over the course of the simulation to ${\simeq}0.45$ and $\rho_{i}=\sqrt{\beta_{i}}d_{i}$
changes modestly. 
The domain resolves scales between $k_{\perp0}\rho_{i0}\equiv2\pi\rho_{i0}/L_{\perp}\approx 0.05$ and $k_{\perp{\rm max}}\rho_{i0}\approx10$, where $\rho_{i0}=\rho_{i}(t=0)$.

\subsection{Decreasing imbalance with distance from the Sun}

A novel feature of this work is the forcing, which changes from imbalanced to balanced over the simulation, heuristically
mimicking the radial evolution of the solar wind. 
We inject energy and cross helicity at the rates $\varepsilon$ and $\varepsilon_{H}$, respectively, with 
 forcing functions $\bm{F}_{\perp}^{U}$ and $\bm{F}_{\perp}^{B}$ that are  
 intended to capture the effect of stirring due to turbulent eddies above the box scale. These
consist of random combinations of large-scale Fourier modes with wavenumbers $k_{j}$ satisfying $2\pi/L_{j}\leq k_{j}\leq4\pi/L_{j}$ for $j=\{x,y,z\}$. 
They are computed as $\bm{F}_{\perp}^{U}=f^{U}\bm{F}_{0}$ and $\grad\btimes\bm{F}_{\perp}^{B}=f^{B}\bm{F}_{0}$,
where $\bm{F}_{0}$ is divergence-free, perpendicular to $\bm{B}_{0}$, and evolved in time 
via an Ornstein--Uhlenbeck process with correlation time $\ta/2$. We fix $\varepsilon$ and $\varepsilon_{H}$ at each time step by adjusting $f^{U}$ and $f^{B}$ 
so that  $n\bm{u}\bcdot\bm{F}_{\perp}^{U}$ and $\bm{B}\bcdot\grad\btimes\bm{F}_{\perp}^{B}$ take the  values needed to inject the required energies into $\bm{z}^{\pm}$.
$\bm{F}_{\perp}^{B}$ is computed from $\bm{F}_{0}$ via a fast Fourier transform and used in the  standard \texttt{Pegasus++}  constrained-transport algorithm
to evolve $\bm{B}$, ensuring that $\grad\bcdot\bm{B}=0$ to machine precision. 
 
\begin{figure}
\includegraphics[width=1.0\columnwidth]{\ffold 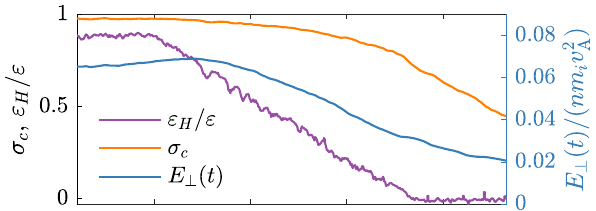}\vspace{-0.06cm}\\
\includegraphics[width=0.865\columnwidth]{\ffold 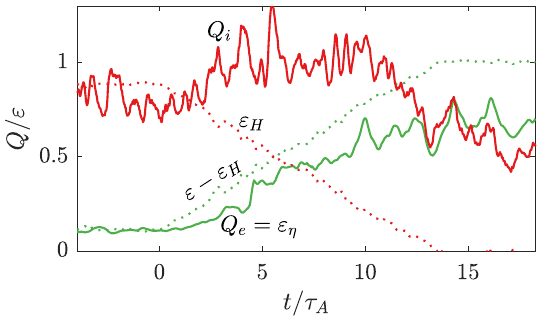}
\caption{({\it Top}) Time evolution of $\varepsilon_H/\varepsilon$ (purple), energy imbalance $\sigma_{c}$ (orange), and total fluctuation energy per unit volume $E_{\perp}(t)$ (blue). ({\it Bottom}) 
Electron heating rate $Q_{e}=\varepsilon_{\eta}$ (green) and ion heating rate $Q_{i}$ (red), with simple helicity-barrier expectations based on the forcing shown with dotted lines. }
\label{fig: time evolution}
\end{figure}

We fix  $\varepsilon=C_{\rm A}m_{i}n(L_{\perp}/L_{z})^{2}\va^{2}V/\ta\approx37m_{i}n\va^{2}\Omega_{i}d_i^3$, where $C_{\rm A}=0.29$ is
the Kolmogorov constant and $V$ is the simulation volume; the factor $(L_{\perp}/L_{z})^{2}\va^{2}/\ta$ guarantees critically balanced fluctuations at the outer scale with $u_{\rm rms}\sim(L_{\perp}/L_{z})\va$ and $\tau_{\rm turb}\sim\ta$.
In contrast, $\varepsilon_{H}$ decreases in time during the
simulation so that the ``injection imbalance'' $\varepsilon_{H}/\varepsilon$ starts at $0.9$ (as in \nathb) then decreases linearly in time at a rate of $0.1$ every $1.5\ta$,
reaching $\varepsilon_{H}=0$ (balanced forcing) at $t=13.5\ta$, where it remains for the rest of the simulation (see \cref{fig: time evolution}). Because the cross helicity and energy are approximately conserved at $k_{\perp}\rho_{i}\ll1$ scales, this evolution is intended to mimic crudely the effect of larger scales becoming 
more balanced with radius, thereby driving the smaller scales with decreasing $\varepsilon_{H}/\varepsilon$.
However, in the actual solar wind, the timescale over which the imbalance decreases is comparable to both the turbulent decay timescale and the expansion timescale $\tau_{\rm exp}\approx R/U$ \citep{Roberts1987,Meyrand2023}. 
By keeping $\varepsilon$ fixed, we  effectively assume that the turbulent decay and expansion are slow compared to the simulation's duration, 
which is appropriate since $\tau_{\rm exp}/\ta\simeq1300\,(B/80~{\rm nT})(R/35~{\rm R}_{\odot})(U/350~{\rm km~s}^{-1})^{-1}$ (see e.g., \citealt{Bale2019}). However, this also implies that our adopted imbalance-decrease timescale of ${\approx}15\ta$ is too short. 
This trade-off is unavoidable given the extreme computational expense of kinetic-turbulence simulations, 
but it should be kept mind that the simulation can only qualitatively capture realistic features of this transition in the 
solar wind.

\section{Results}\label{sec: results}

\begin{figure}
\begin{center}
\includegraphics[width=1.0\columnwidth]{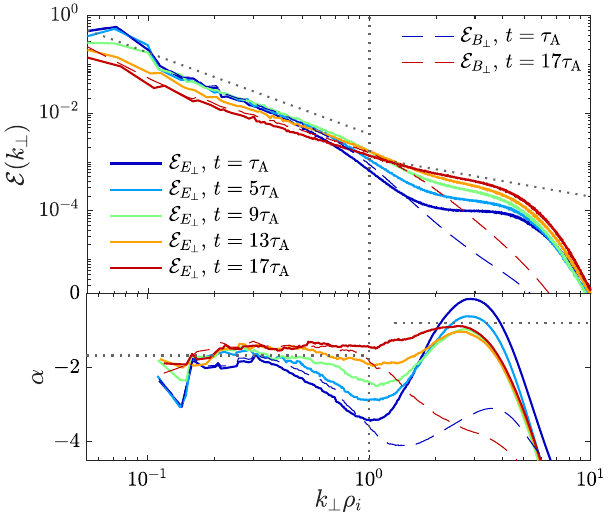}\\
~~\includegraphics[width=0.96\columnwidth]{\ffold 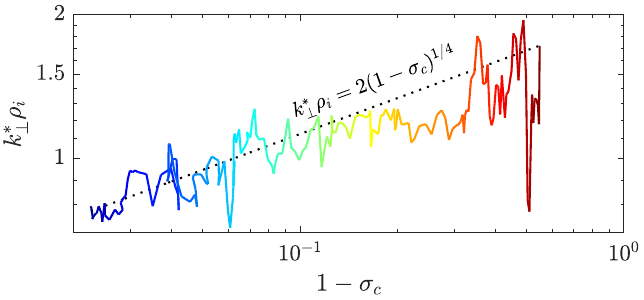}
\caption{(\emph{Top}) Perpendicular spectra of the electric field ($\mathcal{E}_{E_\perp}$; solid lines) and magnetic field ($\mathcal{E}_{B_\perp}$; dashed lines) at 
a selection of times. The sub-panel illustrates the local spectral slopes $\alpha$, computed via a fit over range $\pm\log_{10}(k_{\perp})\approx0.1$ around each point; the dotted lines indicate $k_{\perp}\rho_{i}=1$ and the representative power laws $k_{\perp}^{-5/3}$ and $k_{\perp}^{-0.8}$. 
(\emph{Bottom}) Evolution of the break scale $k_{\perp}^{*}$ with imbalance $1-\sigma_{c}$, 
with the line color indicating the time  as in the top 
panel. The dotted line shows the empirical scaling $k_{\perp}^{*}\rho_{i}=2(1-\sigma_{c})^{1/4}$, which
provides a good fit to the simulation until $t\approx11\ta$.}
\label{fig: spectra}
\end{center}
\end{figure}

\subsection{Evolution of the helicity barrier}

The time evolution as the injection imbalance decreases is shown in \cref{fig: time evolution}. Properties of the saturated imbalanced turbulence, from which the system is initialized, are discussed  in \nathb; a 
portion of this phase is shown at $t<0$ for comparison. The  imbalance of the 
fluctuations,
\begin{equation}
\sigma_{c}\equiv\frac{\langle|\bm{z}^{+}|^{2}\rangle-\langle|\bm{z}^{-}|^{2}\rangle}{\langle|\bm{z}^{+}|^{2}\rangle+\langle|\bm{z}^{-}|^{2}\rangle}=\frac{2\langle\bm{u}_{\perp}\bcdot\bm{b}_{\perp}\rangle}{\langle|\bm{u}_{\perp}|^{2}\rangle+\langle|\bm{b}_{\perp}|^{2}\rangle},
\end{equation}
starts at $\sigma_{c}\approx0.98$ for $t<0$, then 
decreases as the system adjusts to the changing forcing ($\varepsilon_{H}/\varepsilon$).
We halt the simulation once $\sigma_{c}\simeq0.4$, a value similar to that in the solar wind around $1~{\rm au}$; as shown below, the heating properties have changed significantly by this point even though $\sigma_{c}$ is still nonzero.  The blue line shows the  
fluctuation energy density $E_{\perp}=\langle nm_{i}(|\bm{u}_{\perp}|^{2}+|\bm{b}_{\perp}|^{2})\rangle/2$,
which decreases over the simulation, as expected because the turbulence can dissipate more effectively  
at lower $\varepsilon_{H}/\varepsilon$.  

The lower panel of \cref{fig: time evolution} 
compares the measured heating rates, $Q_{i}$ and $\varepsilon_{\eta}$, to helicity-barrier predictions. Here $\varepsilon_{\eta}$, the  hyper-resistive dissipation,
is taken as a proxy for the electron-heating rate $Q_{e}$, which assumes that ions cannot interact with $k_{\perp}\rho_{i}\gg1$ eddies and that there is minimal electron heating via ICWs or other  $k_{\perp}\rho_{i}\lesssim1$ fields \citep{Chandran2010a}.
For most of the simulation, $\varepsilon_{\eta}$ (green line) tracks well the helicity-barrier prediction $Q_{e}=\varepsilon-\varepsilon_{H}$ (dotted-green line),
aside from a delay of ${\sim}\ta$ 
as the injected energy cascades towards smaller scales. 
$Q_i$ increases initially because the energy $E_{\perp}(t)$ decreases while the sub-$\rho_{i}$ energy flux is fixed by the
helicity barrier. Its large value results from the unrealistically fast decrease of $\varepsilon_{H}/\varepsilon$: in reality
 $E_{\perp}(t)$ would change more slowly, implying that $Q_{i}=\varepsilon-Q_e-\partial_{t}E_{\perp}\approx\varepsilon_{H}-\partial_{t}E_{\perp}$  should more closely track $\varepsilon_{H}$.
 As $\varepsilon_{H}/\varepsilon$ nears zero ($t/\tau_{\rm A}\gtrsim 10$) and the helicity barrier erodes, the heating departs from $Q_e \approx\varepsilon-\varepsilon_{H}$ and approaches that found in kinetic simulations of balanced turbulence at $\beta\approx0.3$ \citep{Kawazura2019,Cerri2021}, in which ion heating via Landau damping and stochastic heating absorbs a portion of the energy flux before it reaches the smallest scales. At around the same time, $Q_{i}$ drops significantly.

\Cref{fig: spectra} shows the evolution of the perpendicular spectra $\mathcal{E}(k_\perp)$ of $\bm{E}_{\perp}$ and $\bm{B}_\perp$. The flatter sub-$\rho_{i}$ range in the former helps to highlight the double-kinked ``transition-range''
power law; for $k_{\perp}\rho_{i}\lesssim 1.5$, including in the transition range, $\mathcal{E}_{E_\perp}\approx\mathcal{E}_{B_\perp}$. 
As  the turbulence becomes balanced, the spectral break smoothly moves towards smaller scales, 
creating a less pronounced transition range that is narrower in $k_{\perp}$ and less steep, eventually transitioning directly into the kinetic range with $\mathcal{E}_{E_\perp}\sim k_{\perp}^{-0.8}$.
At late times, $\delta B_{\perp}$ is resistively damped by the
larger resistivity and the magnetic spectrum does not 
exhibit a clear ${\sim}k_{\perp}^{-2.8}$ kinetic  range.

The lower panel of \cref{fig: spectra} provides the evolution of the break scale $k^{*}_\perp$, obtained by fitting a broken power-law to $\mathcal{E}_{B_\perp}(k_{\perp})$.\footnote{
The functional form of the fit is $\mathcal{E}_{\rm fit}=[(k_{\perp}/k_{\perp}^{*})^{n\alpha_{1}}+(k_{\perp}/k_{\perp}^{*})^{n\alpha_{2}}]^{-1/n}$, where $\alpha_1$ ($\alpha_2$) is the power-law index for $k_\perp<k_{\perp}^{*}$ ($k_\perp>k_{\perp}^{*}$), and $n$ controls the break's sharpness; we fit $\mathcal{E}_{B_\perp}(k_\perp)$ in the range $0.14<k_{\perp}\rho_{i}<2$. The measured proportionality between $k_{\perp}^{*}\rho_i$ and $(1-\sigma_{c})^{1/4}$ varies with fitting choices, but the power-law exponent ($1/4$) is robust.} We find a clear power-law dependence on the energy imbalance, $k_{\perp}^{*}\rho_{i}\propto(1-\sigma_{c})^{1/4}\sim(z^{-}_{\rm rms}/z^{+}_{\rm rms})^{1/2}$, for $\sigma_{c}\gtrsim0.8$ ($t\lesssim11\ta$) when the barrier is active ($\varepsilon_{H}/\varepsilon\gtrsim0$).
Although this scaling remains unexplained theoretically, it matches the results of low-$\beta$ gyrokinetic simulations for various choices of $\varepsilon_{H}/\varepsilon$ (\citealt{Meyrand2021}, figure 7), suggesting
it is a robust consequence of the helicity barrier and independent of the energy-dissipation mechanism (gyrokinetics is ignorant of ICW kinetic physics). The 
persistence of the transition-range drop, as well as the evolution of $\varepsilon_{\eta}$ and $Q_{i}$ in \cref{fig: spectra}, provide good evidence that a helicity barrier actively mediates the turbulence 
until $t\approx11\ta$, at which point it transitions towards the balanced regime.

\begin{figure*}
\begin{center}
\includegraphics[width=1\textwidth]{\ffold 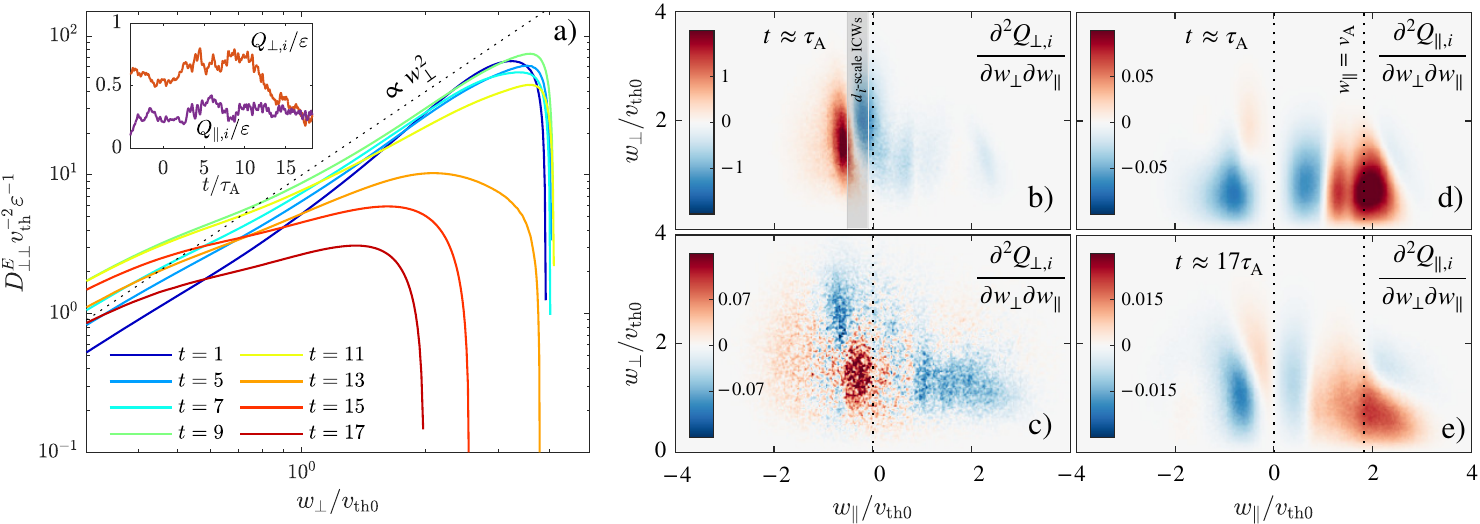}
\caption{(\emph{Left}) Time evolution of the perpendicular diffusion energy coefficient $D_{\perp\perp}^{E}$, showing the transition from the quasi-linear expectation $D_{\perp\perp}^{E}\propto w_{\perp}^{2}$ to a flatter profile around $t\gtrsim 11\ta$. The inset shows the normalized perpendicular (orange) and parallel (purple) ion heating rates, illustrating a sharp drop in $Q_{\perp,i}$ associated with the change in $D_{\perp\perp}^{E}$, while $Q_{\|,i}$ remains almost constant.
(\emph{Right}) Differential heating rates, $v_{\rm th0}^{2}\varepsilon^{-1}\partial^{2}Q_{\perp,\|,i}/\partial w_{\perp}\partial w_{\|}$, computed from $\langle\bb{E}_{\perp,\|}\bcdot\bb{w}_{\perp,\|}\rangle$. The  $Q_{\perp}$ profile (panels~b~and~c) exhibits resonant structure at early times (top); the shaded region shows $w_{\|{\rm res}}$ for oblique
ICWs with $d_{i}^{-1}<k_{\|}<2d_{i}^{-1}$, around where there is a sharp dropoff in wave power (\cref{fig: ICWs 2D}). By late times it drops significantly and an ill-defined peak around $w_{\perp}\simeq v_{\rm th}$ appears, reminiscent of  stochastic heating. The $Q_{\|}$ profile (panels~d~and~e) maintains its structure, although spreads out modestly (accounting partially for its lower values).
}
\label{fig: ion heating}
\end{center}
\end{figure*}

\begin{figure*}
\begin{center}
\includegraphics[width=1.0\textwidth]{\ffold 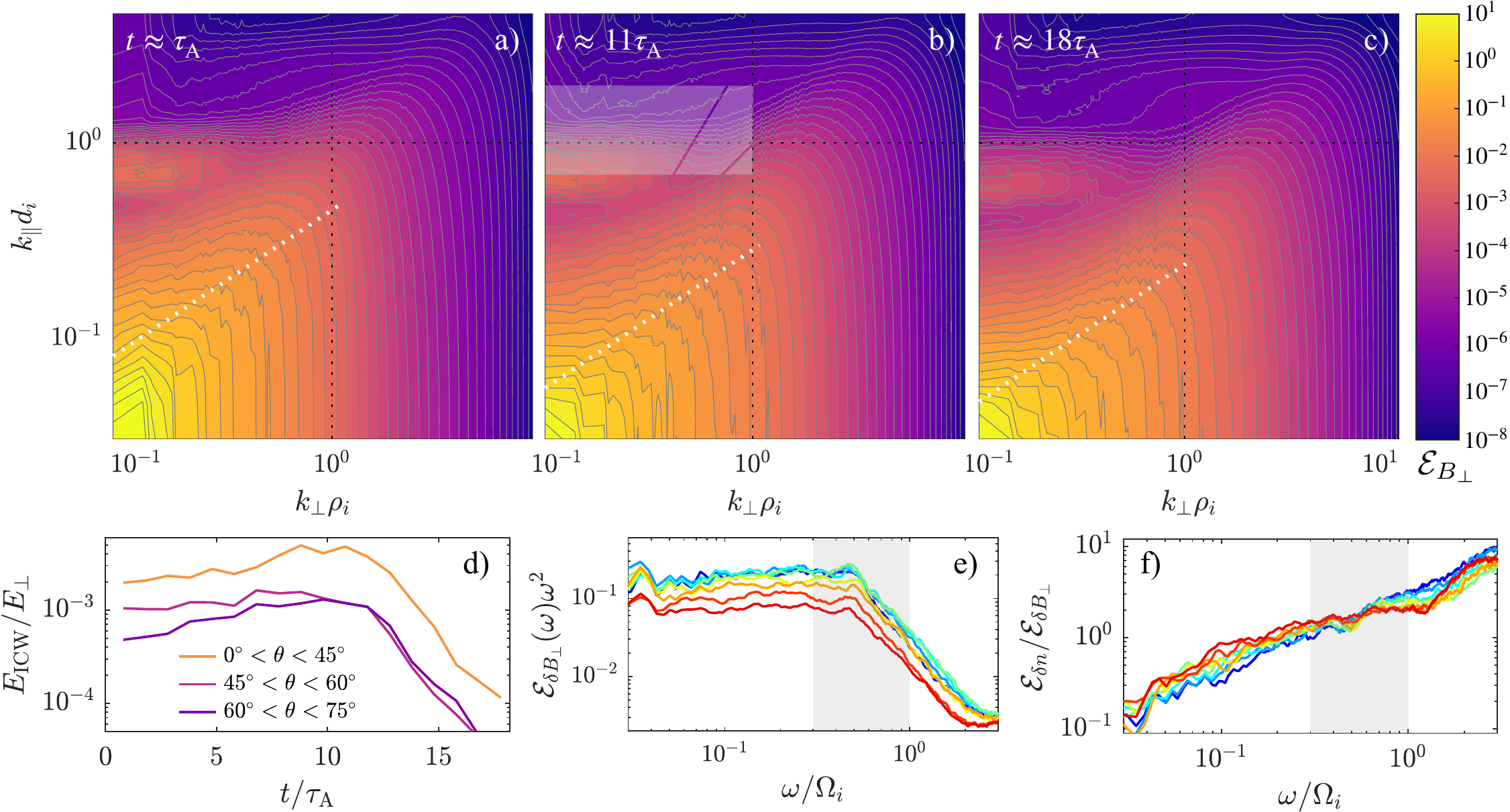}\caption{(a)--(c) 2D spectra of $\delta B_\perp$, $\mathcal{E}_{B_\perp}(k_{\perp},k_{\|})$, at $t\approx\ta$ (left), $t\approx11\ta$ (around the time of the heating transition; middle), and $t\approx18\ta$ (right).  The dotted white lines show the critical-balance condition $k_{\|}\va=2 k_{\perp}z^{+}_{\rm rms}(k_{\perp}/k_{\perp0})^{-1/3}$.
(d) $E_{\rm ICW}$ for different ICW obliquities, which is the energy contained in the shaded regions of panel~(b) (see text). (e)--(f) 
Frequency spectra at different times, using the same colors as \cref{fig: ion heating}a, showing (e) the magnetic-field spectrum $\omega^2\mathcal{E}_{\delta B_{\perp}}(\omega)$  and (f) the 
ratio of density to magnetic-field spectra $\mathcal{E}_{\delta n}/\mathcal{E}_{\delta B_{\perp}}$.
Because oblique ICWs involve $\delta n$ fluctuations, 
while parallel ICWs do not, the drop in the ratio $\mathcal{E}_{\delta n}/\mathcal{E}_{\delta B_{\perp}}$ with time at high frequencies is taken as a signature of a drop in the 
relative power in oblique ICWs compared to parallel ICWs. }
\label{fig: ICWs 2D}
\end{center}
\end{figure*}

\subsection{Consequences for ion heating}

This  drop in ion heating associated with the balanced transition is diagnosed in \cref{fig: ion heating}. We show the perpendicular energy-diffusion coefficient $D_{\perp\perp}^{E}$ (panel~a), which is computed from the measured evolution of $f_{i}$ via
\begin{equation}
D_{\perp\perp}^{E}=\left(\frac{\partial f_{i}}{\partial e_{\perp}}\right)^{-1}\int_{0}^{e_{\perp}}{\rm d}e_{\perp}'\frac{\partial f_{i}(e_{\perp}')}{\partial t},\label{eq: Dpp computation}
\end{equation}
where $e_{\perp}\equiv w_{\perp}^{2}/2$ and $f_{i}(e_{\perp})=\int{\rm d}w_{\|}w_{\perp}\,f_{i}(w_{\perp},w_{\|})$. Equation~\eqref{eq: Dpp computation} 
is taken directly from the $D_{\perp\perp}^{E}$ definition, $\partial f_{i}/\partial t=\partial/\partial e_{\perp}(D_{\perp\perp}^{E}\partial f_{i}/\partial e_{\perp})$ \citep{Vasquez2020},
averaging over the $w_{\|}$ variation of $f_{i}$ and assuming that the heating is predominantly perpendicular, as appropriate for $t\lesssim12\ta$ (\cref{fig: ion heating}a inset).
In quasi-linear theory, waves strongly scatter
``resonant'' particles with
$w_\|=w_{\|{\rm res}}\equiv\omega(\bm{k})/k_\|-n\Omega_i/k_\|$ for $n\in\mathbb{Z}$, flattening $f_{i}$ along contours of constant energy in the wave's frame (the ``resonance contours''; we consider only $n=1$ resonances). 
When parallel waves dominate the spectrum,  $D_{\perp\perp}^{E}\propto w_{\perp}^{2}$, because $w_{\|{\rm res}}$ and various wave-intensity/polarization 
factors are all independent of $w_{\perp}$~\citep{Kennel1966}.
Although the wave spectrum here is more complex, involving a mix of oblique and parallel ICW modes (see below),
this prediction is nonetheless well satisfied until $t\approx 11\ta$.\footnote{
The effect of obliquity is to flatten $D_{\perp\perp}^{E}$ at larger $w_{\perp}$, particularly for shorter-wavelength waves \citep{Isenberg2011}. The highest-$k_\|$ power here is predominantly parallel  (\cref{fig: ICWs 2D}) so these effects are likely minor.}  
After this, $D_{\perp\perp}^{E}$ drops 
and flattens, causing a large drop in the perpendicular ion heating $Q_{\perp,i}$, even while 
the parallel heating $Q_{\|,i}$ remains almost constant (see inset). At  later times, the form of $D_{\perp\perp}^{E}$ does not clearly indicate 
a particular heating mechanism,
but is plausibly consistent with stochastic heating \citep{Chandran2010}: computing the $D_{\perp\perp}^{E}$ predicted for stochastic heating using the  method of \citet{Cerri2021} gives a similar shape (not shown). 
As a complementary analysis, panels~(b)~and~(c) display  $\partial^{2}Q_{\perp,i}/\partial w_{\perp}\partial w_{\|}$, the differential perpendicular heating per unit velocity, which is computed directly from $\langle\bm{E}_{\perp}\bcdot\bm{w}_{\perp}\rangle$ evaluated along particle trajectories (the so-called 
field-particle correlation technique; \citealp{Klein2016a,Arzamasskiy2019}). At early times 
a resonant feature is centered around the $w_\|$ that resonates with the smallest-($d_i$)-scale oblique ICWs with significant power (the shaded region shows $w_{\|{\rm res}}$ for $d_{i}^{-1}\lesssim k_{\|}\lesssim2d_{i}^{-1}$; see \cref{fig: ICWs 2D}); at later times, the magnitude drops significantly 
into a diffuse peak around 
$w_{\perp}\simeq v_{\rm th}$, consistent with stochastic heating. In contrast, $\partial^{2}Q_{\|,i}/\partial w_{\perp}\partial w_{\|}$ (the differential parallel heating; panels~d~and~e), maintains a form consistent with Landau damping of kinetic Alfv\'en waves (KAWs) throughout the simulation.

In \cref{fig: ICWs 2D} we provide evidence that the shut off in ion heating occurs because 
the turbulence amplitude decreases to the point where it can no longer drive quasi-linear heating by oblique ICWs.
In the top panels, we show  2D spectra of $\delta B_{\perp}$ fluctuations, $\mathcal{E}_{B_\perp}(k_{\perp},k_{\|})$, which 
are computed by filtering each field into $k_{\perp}$ bins, then interpolating these onto the exact magnetic-field lines to take a $k_{\|}$ spectrum (see Methods in \nathb). Energy is concentrated at $k_{\perp}>k_{\|}$ (the turbulence) and in a bump at $k_{\perp}\ll k_{\|}\simeq0.8d_{i}$ (parallel ICWs). Examining the time evolution from left to right, we see that
the energy migrates to lower $k_{\|}$ with time  (it moves downwards), with the 
 dotted white lines indicating how this follows the critical-balance scaling $k_{\|}\va=Ak_{\perp}z^{+}_{\rm rms}(k_{\perp}/k_{\perp0})^{-1/3}$ 
 (the coefficient $A=2$ is chosen to align with the ``peak'' on the cone and is consistent across time). 
A corollary is that larger-amplitude turbulence feeds more power into smaller-$k_\|$ oblique ICWs with smaller $w_{\|{\rm res}}$, driving more quasi-linear heating. 

\begin{figure}
\begin{center}
~~~~~\includegraphics [width=0.89\columnwidth]{\ffold 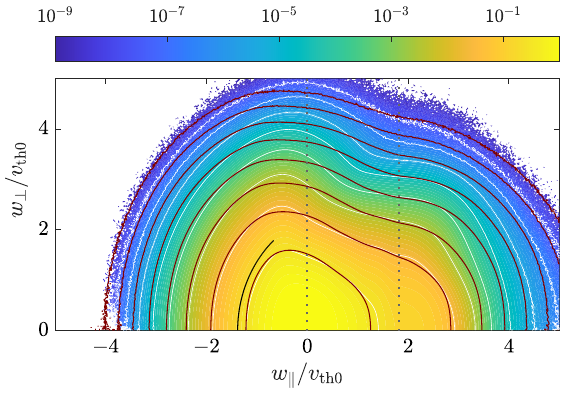}\\
\includegraphics [width=0.95\columnwidth]{\ffold 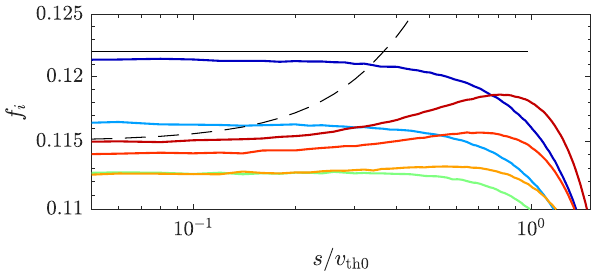}
\caption{(\emph{Top}) Contours of the ion VDF $f_{i}$ at $t\approx5\ta$ (white lines and colormap) and the $t\approx18\ta$ (red lines). Dotted vertical lines  highlight $w_{\|}=0$ and $w_{\|}=\va$. The red contour levels are chosen to emphasize changes in $f_{i}$ and do not correspond to the white contours (see text). (\emph{Bottom}) $f_{i}$ versus distance $s$ along the oblique-ICW resonant contour
shown with the black line in the top panel ($s=0$ at $w_{\|}=0$). Colors are as in  \cref{fig: ion heating}a. The
thin-black line is flat ($\mathcal{G}_{\rm res}[f_{i}]=0$), while the dashed line shows an $f_{i}$ that is independent of $w_{\perp}$ (as 
expected from stochastic heating).}
\label{fig: VDFs}
\end{center}
\end{figure}

We quantify this effect in panel~(d) by computing the  energy of ICWs ($E_{\rm ICW}$), defined as fluctuations with $0.7\leq k_{\|}d_{i}\leq2$, as a function of wavevector obliquity $\theta=\tan^{-1}(k_{\perp}/k_{z})$. The minimum wavenumber $k_{\|}d_{i}=0.7$ is chosen to capture fluctuations with $w_{\|{\rm res}}\lesssim v_{\rm th}$, which interact with the VDF core \citep{Isenberg2011}, while the angle ranges capture quasi-parallel ($0<\theta<45^{\circ}$), moderately oblique ($45^{\circ}<\theta<60^{\circ}$), and highly oblique ($60^{\circ}<\theta<75^{\circ}$) populations; these ranges are indicated by the shaded regions in panel~(b).
Despite $E_\perp$ decreasing continuously, the energies of the oblique-ICW
populations  increase slightly for $t\lesssim12\ta$ before  dropping rapidly. 
The similarity of this functional form with that of  $Q_{\perp,i}(t)$ provides good evidence for oblique ICWs being the primary driver of heating. 
The energy of parallel ICWs ($\theta\lesssim45^{\circ}$) follows a similar trend, but they are not  driven directly by the turbulence, which has little power at $k_{\|}\gtrsim k_{\perp}$. Instead they arise because
oblique-ICW quasi-linear heating causes $f_{i}$ to increase along the resonance contours of parallel ICWs, thus emitting waves (causing 
instability; \citealp{Kennel1967,Chandran2010a}). Their energy is driven and undamped, building up substantially for $t\lesssim12\ta$, then decreasing at later times as  $f_{i}$ changes shape and renders parallel ICWs stable. Further evidence for this 
scenario is presented in panels~(e) and (f), which show frequency spectra $\mathcal{E}(\omega)$ of $\delta B_{\perp}$  and density  fluctuations. These 
are computed from high-cadence time data at $100$ spatial points, segmented into time 
intervals of $\pm2\ta = \pm809\Omega_i^{-1}$ around different times matching \cref{fig: ion heating} (we highlight the ICW frequency range where $k_{\|}d_{i}\gtrsim 0.7$, $0.3\lesssim\omega/\Omega_{i}\lesssim1$). 
In $\mathcal{E}_{\delta B_\perp}(\omega)$, the lower-frequency fluctuations approximately satisfy $\mathcal{E}_{\delta B_{\perp}}\propto \omega^{-2}$, which signifies a constant-flux cascade \citep{Corrsin1963,Beresnyak2015}. The
spectrum drops off dramatically for $\omega\gtrsim 0.5\Omega_i$, which 
is presumably where ICW heating becomes dominant.
In panel~(f) the ratio of density to magnetic fluctuations, $\mathcal{E}_{\delta n}/\mathcal{E}_{\delta B_{\perp}}$, provides a
diagnostic of parallel and oblique ICWs: while both parallel and oblique ICWs involve $\delta B_{\perp}$ fluctuations,
only oblique modes involve $\delta n$.
We see that at  $\omega\gtrsim 0.5\Omega_{i}$, $\mathcal{E}_{\delta n}/\mathcal{E}_{\delta B_{\perp}}$ decreases at intermediate times, even though at low frequencies it increases over the same time period (presumably because larger density fluctuations are driven by the forcing
in balanced turbulence). This high-frequency decrease is consistent 
with the ratio of oblique-ICW to parallel-ICW power seen in 
panel~(d), which drops by a factor of ${\simeq}2$ between $t\approx 5\ta$ and $t\approx 10\ta$ then remains approximately constant over the rest of the simulation.


The interpretation described above predicts that $Q_{i}$ scales with the oblique-ICW power, which itself scales with $k_{\|,{\rm max}}d_{i}$, where $k_{\|,{\rm max}}$ corresponds to the smallest parallel scale accessible by the imbalanced portion of the cascade. An independent estimate of $k_{\|,{\rm max}}d_{i}$ can be obtained from the spectrum using $k_{\|,{\rm max}}\va\sim k_{\perp}^{*}z^{+}_{\rm rms}(k_{\perp}^{*}/k_{\perp0})^{-1/3}$, where $k_{\perp}^{*}\rho_{i}\propto(z^{-}_{\rm rms}/z^{+}_{\rm rms})^{1/2}$ is the break scale (\cref{fig: spectra}). This gives
$k_{\|,{\rm max}}d_{i}\propto\beta^{-1/2}(k_{\perp0}\rho_{i})^{1/3}(z^{+}_{\rm rms})^{2/3}(z^{-}_{\rm rms})^{1/3}/\va$.  In the simulation, $(z^{+}_{\rm rms})^{2}z^{-}_{\rm rms}$ increases modestly  for $t\lesssim11\ta$ before dropping significantly (not shown), remaining well correlated with $Q_{i}$ and $E_{\rm ICW}$ at all times, including during pre-saturated phase of \nathb. This agreement demonstrates the internal consistency of the results across diverse diagnostics ($E_{\rm ICW}$,  $Q_{i}$, and $k_{\perp}^{*}$ via $\mathcal{E}_{B_{\perp}}(k_{\perp})$), and could prove useful for developing a  closure model for helicity-barrier-mediated turbulence.

The effect of these dynamics on $f_{i}$ is illustrated in \cref{fig: VDFs}, whose top panel superimposes isocontours 
of $f_{i}(w_{\perp},w_{\|})$ at $t\approx18\ta$ over $f_i(w_{\perp},w_{\|})$ at $t\approx5\ta$.
Particles are scattered to render $f_{i}$ constant along the resonant contours, which are computed from $\mathcal{G}_{\rm res}[f_{i}]=0$, where $\mathcal{G}_{\rm res}\equiv (1-w_{\|}/v_{\rm ph})\partial/\partial w_{\perp}+(w_{\perp}/v_{\rm ph})\partial/\partial w_{\|}$ with $v_{\rm ph}(w_{\|})=\omega /k_{\|}$ for resonant waves that satisfy $\omega/k_{\|}-\Omega_{i}/k_{\|}=w_{\|}$. With heating 
being driven by oblique ICWs but exciting parallel ICWs for $t\lesssim11\ta$,  we expect $f_{i}$ to decrease along 
the oblique-ICW contours  (which approximately satisfy the $\sin\theta\approx1$ cold-plasma relation $\omega=k_{\|}\va/\sqrt{1+k_{\|}^{2}d_{i}^{2}}$; \citealp{Stix1992}) and to increase along the parallel-ICW contours
($\omega=k_{\|}\va\sqrt{1-\omega/\Omega_{i}}$; see \citealt{Isenberg2012} for further discussion of cold-plasma relations in this context). Its evolution towards a flatter core at $t\gtrsim11\ta$ is hard to 
discern on $f_{i}(w_{\perp},w_{\|})$, so in the lower panel we plot $f_{i}$ along an oblique-ICW resonant contour (thin black line).
It  changes from decreasing slightly to increasing along the contour, with the latter a signature of $f_{i}(w_{\perp},w_{\|})$ becoming flatter in $w_{\perp}$, likely due to stochastic heating (a $w_{\perp}$-independent $f_{i}$ is shown with the dashed line;
\citealp{Klein2016,Cerri2021}). Various other modifications to $f_{i}$ over the simulation are clear in the top panel, including
 the smoothing of the sharp ``ridge'' bordering the quasi-linearly heated region of phase space at $w_{\|}/v_{\rm th0}\approx-0.5$, and 
 the flattening of $f_{i}(w_{\perp})$ at $w_{\|}>0$.

Another feature of $f_{i}$ is the strong parallel plateau, or beam, which results from the Landau damping of $k_{\perp}\gg k_{\|}$  fluctuations at $k_{\perp}\rho_{i}\lesssim1$  (see also \nathb; \citealt{Li2010}). 
Such fluctuations, which lie in the $k_{\perp}$ range between Alfv\'en waves and KAWs, propagate with phase speed ${>}\va$ \citep{Howes2006}, 
growing a modestly super-Alfv\'enic plateau in $f_{i}$. As the turbulence becomes  balanced, such 
fluctuations would likely flatten $f_{i}$ at $w_{\|}\approx-\va$ also (see figure~7 of \citealt{Arzamasskiy2019}),
but this feature is not yet observable. The total parallel heating $Q_{\|,i}$, which captures this beam formation, remains almost  constant throughout the full simulation (\cref{fig: ion heating}a inset).
 This suggests that the $k_{\perp}\rho_{i}\sim1$ Alfv\'enic 
fluctuations are Landau damping some fixed portion of their energy  before dissipating into oblique ICWs (at earlier times) or KAW turbulence (at later times). A corollary is that there is no fundamental difference between beam formation and standard resonant parallel heating, aside from the fluctuations' imbalance.

\section{Discussion}\label{sec: discussion}

A complete theory for the helicity barrier would be able to predict the parameters ($\varepsilon_{H}/\varepsilon$, $\beta$, etc.) at which it is operative. Low-$\beta$ gyrokinetics states only that the barrier occurs unless 
the generalized helicity can be destroyed at least as fast as ${\sim}\varepsilon_H$ before  $k_\perp\rho_i\sim1$ scales are reached \citep{Meyrand2021}. Although generalized helicity is not a true invariant of hybrid kinetics, we can infer from our simulation, based on the agreement between $Q_{e}$ and $\varepsilon-\varepsilon_{H}$  (\cref{fig: time evolution}) and the continuous evolution of the transition-range spectrum (\cref{fig: spectra}) for 
$t\lesssim 11\ta$, that the critical $\varepsilon_{H}/\varepsilon$ at $\beta\approx 0.4$ is   ${\approx}0.2$.
The changes that occur once $\varepsilon_{H}/\varepsilon\lesssim0.2$ also coincide with the sharp drop in $Q_{\perp,i}$ (\cref{fig: ion heating}), suggesting that breaking the helicity barrier curtails ion heating.  These results highlight the surprising robustness of the helicity barrier in the face of the additional complexities of a true kinetic system. 
However, $\varepsilon_H/\varepsilon$ decreases unrealistically fast in our simulation; understanding how the barrier evolves when $\varepsilon_{H}/\varepsilon$ changes on   timescales comparable to the turbulent decay time should be a priority for future work.

Our results provide a helpful roadmap for understanding the radially dependent interplay between turbulence, 
heating, and instabilities in the $\beta\lesssim1$ solar wind.
At smaller $R$ and in faster streams, which have higher observed imbalance, we predict strong perpendicular ion heating  via quasi-linear resonance, continual emission (instability)
of parallel ICWs, a steep and wide ion-Larmor-scale transition range, proton beam formation, and little electron heating. At larger $R$ and in slower streams, which have lower imbalance ($\sigma_{c}\lesssim 0.8$
in our simulation), 
electron heating dominates, parallel ICWs are absorbed/damped,  there is no transition-range spectrum, and sharp features in the VDF are smoothed out. 
These correlations and features match those measured in the low-$\beta$ solar wind by \emph{in situ} spacecraft \citep[e.g.,][]{Marsch2006,Bruno2014,Zhao2021,Shi2023}. 

Under the assumption that, far from the Sun, faster wind is heated
more than slower wind \citep{Hansteen1995,Totten1995,Halekas2023}, our results  explain qualitatively  various interesting properties of observed temperature profiles. 
Close to the Sun, proton and minor-ion temperatures correlate positively with wind speed \citep{Burlaga1973}, 
while the electron temperature is negatively correlated \citep{Marsch1989}; this is natural if the highly imbalanced turbulence 
of faster streams has low $Q_e/Q_i$ \citep{Shi2023}. At larger distances (${\sim}1~{\rm au}$), the electron-temperature correlation flips to positive for $U\lesssim500~{\rm km~s}^{-1}$ \citep{Shi2023}, as would  occur  if $Q_e/Q_i$ increased to ${\gtrsim}1$ as the turbulence becomes balanced at larger radii. More directly, \citet{Abraham2022} measure  $Q_e\propto R^{-2}$ for $R\lesssim0.3~{\rm au}$, followed by a rapid drop in $Q_e$ at larger $R$. This profile, $Q_e\propto R^{-2}$, is much flatter than the ``standard'' total-heating profile $Q\propto R^{-4}$ \cite[e.g.,][]{Totten1995}---a natural 
explanation is that $Q=Q_i+Q_e$ is steeper than $Q_e$, because $Q_e/Q_i$ increases as $\varepsilon_H/\varepsilon$ decreases with $R$. Then, once $\varepsilon_H/\varepsilon\ll1$ (around $R\simeq0.3$ in this scenario),  $Q_e/Q_i$ saturates and $Q_e$ drops rapidly. 

Together, these observations suggest that imbalanced turbulence and the helicity barrier are actively shaping global coronal and solar-wind dynamics. More generally, they highlight the crucial role of imbalance in controlling collisionless plasma thermodynamics. 

\vspace{0.3cm}
\noindent We thank S.~S.~Cerri, C.~H.~K.~Chen, B.~D.~G.~Chandran, E.~Quataert, and A.~A.~Schekochihin for useful discussions.
JS and RM acknowledge the support of the Royal Society Te Ap\=arangi, through Marsden-Fund grants MFP-UOO2221 (JS) and MFP U0020 (RM), as well as through the Rutherford Discovery Fellowship  RDF-U001804  (JS).
This research was part of the Frontera computing project
 at the Texas Advanced Computing Center, which is made possible by National Science Foundation award OAC-1818253. Further computational support was provided by the New Zealand
 eScience Infrastructure (NeSI) high performance computing facilities, funded jointly by NeSI’s
 collaborator institutions and through the NZ MBIE, and through PICSciE-OIT TIGRESS High
 Performance Computing Center and Visualization Laboratory at Princeton University.

\software{{\tt Pegasus++ }}


\end{document}